\documentclass[aps,prd,preprint,showpacs,groupedaddress,floatfix]{revtex4-1}
\usepackage{lineno}
\modulolinenumbers[5]

\usepackage{graphicx}
\usepackage{dcolumn}
\usepackage{bm}
\usepackage{amsmath}
\usepackage{amssymb}
\usepackage{makecell}
\usepackage{array}
\usepackage[utf8]{inputenc}
\usepackage{cancel}
\usepackage{subcaption}

\newcommand{\beq}{\begin{equation}}
\newcommand{\eeq}{\end{equation}}
\newcommand{\bea}{\begin{eqnarray}}
\newcommand{\eea}{\end{eqnarray}}

\begin{document}

\title{Flux tubes in the QCD vacuum}

\author{Paolo Cea}
\email{paolo.cea@ba.infn.it}
\affiliation{Dipartimento di Fisica dell'Universit\`a di Bari, I-70126 Bari, 
Italy \\
and INFN - Sezione di Bari, I-70126 Bari, Italy}

\author{Leonardo Cosmai}
\email{leonardo.cosmai@ba.infn.it}
\affiliation{INFN - Sezione di Bari, I-70126 Bari, Italy}

\author{Francesca Cuteri}
\email{lcuteri@th.physik.uni-frankfurt.de}
\affiliation{Institut f\"ur Theoretische Physik, Goethe Universit\"at,
            60438 Frankfurt am Main, Germany}

\author{Alessandro Papa}
\email{papa@cs.infn.it}
\affiliation{Dipartimento di Fisica dell'Universit\`a della Calabria,
I-87036 Arcavacata di Rende, Cosenza, Italy \\
and INFN - Gruppo collegato di Cosenza, I-87036 Arcavacata di Rende, Cosenza, 
Italy}

\date{\today}           

\begin{abstract}
The hypothesis that the QCD vacuum can be modeled as a dual superconductor is a
powerful tool to describe the distribution of the color field
generated by a quark-antiquark static pair and, as such, can provide
useful clues for the understanding of confinement.
In this work we investigate, by lattice Monte Carlo simulations of the $SU(3)$ 
pure gauge theory and of (2+1)-flavor QCD with physical mass settings,
some properties of the chromoelectric flux tube at zero temperature
and their dependence on the physical distance between the static sources.
We draw some conclusions about the validity domain of the dual superconductor
picture.
\end{abstract}
  
\pacs{11.15.Ha, 12.38.Aw}

\maketitle


\section{Introduction}
\label{sect:intro}
The confinement of quarks and gluons inside hadrons is a well established
experimental fact, but a theoretical explanation of the underlying dynamics
within the theory of strong interactions, Quantum ChromoDynamics (QCD), is
still missing. Numerical simulations by Monte Carlo methods of QCD
on a space-time lattice provide us with a powerful nonperturbative tool
to probe the vacuum structure of the theory and can help us in
catching some relevant information at the basis of the confinement
phenomenon.

One well established fact, ascertained by a wealth of numerical analyses
in QCD, is that the chromoelectric field between two static quarks
distributes in tubelike structures or ``flux tubes''~\cite{Fukugita:1983du,Kiskis:1984ru,Flower:1985gs,Wosiek:1987kx,DiGiacomo:1989yp,DiGiacomo:1990hc,Singh:1993jj,Cea:1992sd,Matsubara:1993nq,Cea:1992vx,Cea:1993pi,Cea:1994ed,Cea:1994aj,Cea:1995zt,Bali:1994de,Haymaker:2005py,D'Alessandro:2006ug,Cardaci:2010tb,Cea:2012qw,Cea:2013oba,Cea:2014uja,Cea:2014hma,Cardoso:2013lla,Caselle:2014eka}. From these tubelike structures a linear
potential between static color charges naturally arises, thus representing
a numerical evidence of color
confinement~\cite{Bander:1980mu,Greensite:2003bk}. \\

As for a possible dynamical mechanism for confinement, long ago 
't Hooft~\cite{'tHooft:1976ep} and Mandelstam~\cite{Mandelstam:1974pi}
conjectured that the vacuum of QCD could behave as a coherent state of color
magnetic monopoles or, in more modern terms, as a dual
superconductor~\cite{Ripka:2003vv,Kondo:2014sta}: the condensation of
color magnetic monopoles would play in the QCD vacuum the same role as
the condensation of Cooper pairs in a standard superconductor.
Indeed, there is a lot of numerical evidence in favor of color magnetic
condensation~\cite{Shiba:1994db,Arasaki:1996sm,Cea:2000zr,Cea:2001an,DiGiacomo:1999fa,DiGiacomo:1999fb,Carmona:2001ja,Cea:2004ux,D'Alessandro:2010xg,Kato:2014nka}, however it cannot be excluded that color magnetic monopole condensation
is a consequence of the mechanism of color confinement~\cite{'tHooft:2004th},
whose origin could be found in some, so far unknown, dynamical effect.
Still, the dual superconductivity picture of the QCD vacuum can serve
as a very useful phenomenological tool to interpret the vacuum dynamics. 
Many previous studies of our collaboration (or of a part of
it)~\cite{Cea:1992vx,Cea:1993pi,Cea:1994ed,Cea:1994aj,Cea:1995zt,Cardaci:2010tb,Cea:2012qw,Cea:2013oba,Cea:2014uja,Cea:2014hma} have indeed furnished clear
evidence that, at zero temperature, color flux tubes, made up 
almost completely by the longitudinal chromoelectric field directed along the line joining a static quark-antiquark
pair, can be successfully described within the dual superconductivity picture,
both in $SU(2)$ and in $SU(3)$ pure gauge theories. 
In our most recent paper~\cite{Cea:2015wjd} the investigation of the
structure of flux tubes in $SU(3)$ was extended to the case of
nonzero temperature and lead to the result that the flux tube
between two static sources separated by a distance of about 0.76~fm
survives even above the critical temperature $T_c$ of the deconfinement
transition, keeping a more or less constant transverse shape, but
housing in it a weaker and weaker chromoelectric field as the temperature
increases. Such (somewhat surprising) phenomenon could be peculiar of the
only value of the distance between the sources considered in that work and
evidently motivates to extend the analysis to different values of the
distance between the sources. As a matter of fact, a careful study, within the
dual superconductor model, of the dependence of the flux tube shape on the
distance between the color sources has not been carried out so far either
at zero temperature, at least in non-Abelian lattice field theories. In the 
three-dimensional Abelian U(1) lattice gauge theory, instead, such an analysis
has been completed very recently~\cite{Caselle:2016mqu}.

The role of the distance $d$ between the static sources for the distribution
of the color fields and, hence, for the shape of the flux tube has been
emphasized in~\cite{Baker:2015zlm}: at small distances the dual
superconductivity picture is expected to hold, whereas the effective string
theory approach~\cite{Luscher:1980ac,Luscher:1980fr,Luscher:1980iy} is
expected to take over at large distances, the transition regime being
localized around $d=2/\sqrt{\sigma}$. According to the effective string theory 
description, the shape of the flux tube is determined by a fluctuating 
thin string connecting the sources. Within this approach the
quark-antiquark potential and the width of the flux tube have been studied 
numerically in $SU(N)$ gauge theories, both at $T=0$ and at $T<T_c$, in many 
papers~\cite{Caselle:2004er,Caselle:2005xy,Caselle:2006wr,Gliozzi:2010jh,Gliozzi:2010zv,Caselle:2011vk,Caselle:2012rp}. In several other
recent works~\cite{Cardoso:2013lla,Bakry:2011kga,Bakry:2014gea,Bakry:2012eq,Bakry:2015csa,Bakry:2011zz,Bakry:2010zt} also the detailed profile of the color
field distribution near static sources has been analyzed.

The aim of this paper is to assess the validity domain of the dual
superconductivity picture of the QCD vacuum, by confronting its predictions
for some of the parameters determining the shape of the flux tube with
Monte Carlo data, when the distance $d$ between the static sources is varied
in the range 0.76~fm to 1.33~fm, corresponding to the range
$1.6/\sqrt{\sigma}$ to $2.8/\sqrt{\sigma}$. The analysis is performed both
in the $SU(3)$ pure gauge theory and, for the first time ever, also in
(2+1)-flavor QCD with physical quark mass settings. The considered range of
distances is the largest one for which the setup of our numerical analysis
allowed the extraction of physical information out of the statistical noise
and is large enough to include the regime where, according
to~\cite{Baker:2015zlm}, the dual superconductivity hypothesis should fail.

The plan of the paper is the following: in Section~\ref{background} we
recall the theoretical background for the dual superconductivity predictions
and introduce the lattice observables used to extract the field strength
tensor of the static quark-antiquark sources; in Section~\ref{setup}
we illustrate our lattice setup and present the numerical results of
our analysis; finally, in Section~\ref{discussion}, we comment on our
findings.
\section{Theoretical background and lattice observables}
\label{background}
The field configurations generated by a static quark-antiquark pair can be
probed by calculating on the lattice the vacuum expectation value of
the following connected correlation
function~\cite{DiGiacomo:1989yp,DiGiacomo:1990hc,Kuzmenko:2000bq,
DiGiacomo:2000va}:
\begin{equation}
\label{rhoW}
\rho_{W,\mu\nu}^{\rm conn} = \frac{\left\langle {\rm tr}
\left( W L U_P L^{\dagger} \right)  \right\rangle}
              { \left\langle {\rm tr} (W) \right\rangle }
 - \frac{1}{N} \,
\frac{\left\langle {\rm tr} (U_P) {\rm tr} (W)  \right\rangle}
              { \left\langle {\rm tr} (W) \right\rangle } \; .
\end{equation}
Here $U_P=U_{\mu\nu}(x)$ is the plaquette in the $(\mu,\nu)$ plane, connected
to the Wilson loop $W$, lying on the $\hat 4 \hat i$-plane, with $\hat i$
  any fixed spatial direction, by a Schwinger line $L$, and $N$ is the number
of colors.
\begin{figure}[htb] 
\begin{subfigure}{0.5\textwidth}
  \centering
\includegraphics[scale=0.7,clip]{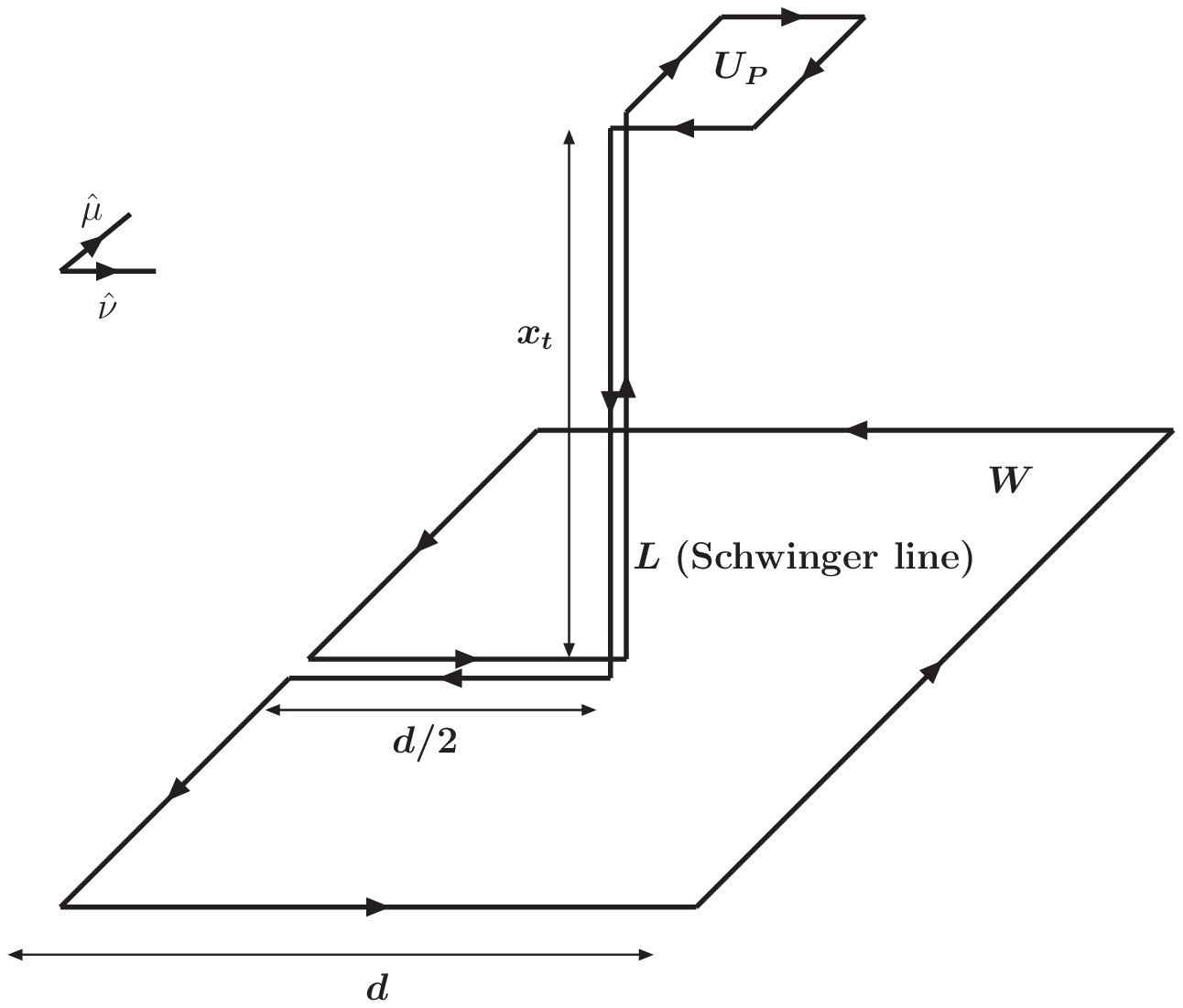} 
\end{subfigure}
\begin{subfigure}{0.3\textwidth}
\includegraphics[scale=0.8,clip]{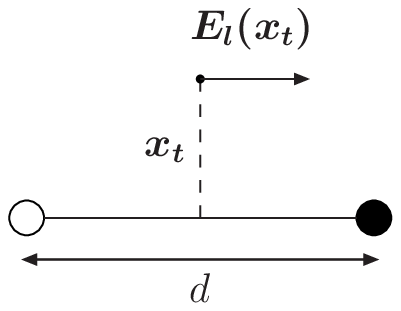}
\end{subfigure}
\caption{(Left) The connected correlator given in Eq.~(\protect\ref{rhoW})
between the plaquette $U_{P}$ and the Wilson loop
(subtraction in $\rho_{W,\mu\nu}^{\rm conn}$ not explicitly drawn).
(Right) The longitudinal chromoelectric field $E_l(x_t)$ with respect to the
position of the static sources (represented by the white and black circles),
for a given value of the transverse distance $x_t$.}
\label{fig:op_W}
\end{figure}
The correlation function defined in Eq.~(\ref{rhoW}) measures the field 
strength, since in the naive continuum limit~\cite{DiGiacomo:1990hc}
\begin{equation}
\label{rhoWlimcont}
\rho_{W,\mu\nu}^{\rm conn}\stackrel{a \rightarrow 0}{\longrightarrow} a^2 g 
\left[ \left\langle
F_{\mu\nu}\right\rangle_{q\bar{q}} - \left\langle F_{\mu\nu}
\right\rangle_0 \right]  \;,
\end{equation}
where $\langle\quad\rangle_{q \bar q}$ denotes the average in the presence of 
a static $q \bar q$ pair and $\langle\quad\rangle_0$ is the vacuum average,
which is expected to vanish. This leads to the following definition of the
quark-antiquark field strength tensor:
\begin{equation}
\label{fieldstrengthW}
F_{\mu\nu}(x) = \frac{1}{a^2 g } \; \rho_{W,\mu\nu}^{\rm conn}(x)   \; .
\end{equation}
In the particular case when the plaquette $U_P$ and the Wilson loop $W$
lie on parallel planes (see Fig.~\ref{fig:op_W} (left) with $\hat \mu=\hat 4$
and $\hat \nu=\hat i$), we get $F_{4i}(x)$, {\it i.e.} the chromoelectric
field in the direction $\hat i$, which is the direction longitudinal to the
axis connecting the two static sources. The position $x$ in space where the
longitudinal chromoelectric field is evaluated depends on the relative
position of the plaquette $U_P$ to the Wilson loop $W$; for the setup
of Fig.~\ref{fig:op_W} (left), the point $x$ is equidistant to the
two static sources and off the axis connecting them (no matter
in which direction, due to the azimuthal symmetry) by a distance $x_t$.
In the following we denote the longitudinal chromoelectric field
$F_{4i}(x)$ by $E_l(x_t)$ (see also Fig.~\ref{fig:op_W} (right)).

In this study we have not considered the effect of changing the path
along which the Schwinger line $L$ is constructed. However, in a
study about gauge-invariant field-strength correlators~\cite{DiGiacomo:2002mq}
the dependence on the shape of the Schwinger line was investigated and it was found that, while
different shapes correspond to differences in the intensity of the
measured field, the slope seemed to be completely path independent.
The same path independence is then plausible for all the physical quantities extracted from our fits.

As far as the color structure of the field $F_{\mu\nu}$ is concerned, we
observe that the Wilson loop connected to the plaquette is the source of
a color field which points, in average, onto an unknown direction $n^a$ in
color space (there is no preferred direction). We thus measure the average
projection of the color field onto that direction. The role of the
Schwinger lines entering the definition~(\ref{rhoW}) is to realize the color
parallel transport between the source loop and the ``probe'' plaquette.
Therefore, the $F_{\mu\nu}$ appearing in Eq.~(\ref{fieldstrengthW}), should be
understood as 
$n^a F_{\mu\nu}^a$,
\begin{equation}
\rho_{W,\mu\nu}^{\rm conn}\stackrel{a \rightarrow 0}{\longrightarrow} a^2 g
\left[ \left\langle
n^aF^a_{\mu\nu}\right\rangle_{q\bar{q}} \right]\;.
\end{equation}
This relation is a necessary consequence of the gauge-invariance of the
operator defined in~(\ref{rhoW}) and of its linear dependence on the
color field in the continuum limit (see Eq.~\eqref{rhoWlimcont}).
An explicit verification of the latter property was exhibited in
Ref.~\cite{Cea:2015wjd} (see Fig.~3 there).

The numerical results presented in this work refer to the longitudinal
chromoelectric field $E_l(x_t)$ for different values of $x_t$ and for
several choices of the distance $d$ between the static sources. 
Due to the azimuthal symmetry, the transverse shape of the longitudinal
chromoelectric field at midway between the static sources can be fully
reconstructed. It is evidently useful to describe this transverse shape
in terms of a few physical parameters, which could possibly help identifying the
underlying mechanism of confinement. The dual superconductor model turns
out to be a powerful tool to describe this transverse shape, at least at
distances $d$ not too large with respect to the inverse square root of the
string tension.

The key assumption of the dual superconductor model is to understand the
chromoelectric flux tube in the QCD vacuum as the dual counterpart of an 
Abrikosov tube inside an ordinary superconductor. According to
this interpretation, the transverse shape of the longitudinal chromoelectric
field $E_l$ should resemble the dual version of the Abrikosov vortex field 
distribution. This naturally leads to the idea of describing chromoelectric
flux tubes by means of the same tube-like solutions of the Ginzburg-Landau
equations in usual electric superconductivity. One such solution was
proposed long ago~\cite{Cea:1992sd,Cea:1992vx,Cea:1993pi,Cea:1994ed,Cea:1994aj,
Cea:1995zt} to fit the transverse shape of the longitudinal chromoelectric 
field:
\begin{equation}
\label{London}
E_l(x_t) = \frac{\phi}{2 \pi} \mu^2 K_0(\mu x_t) \;,\;\;\;\;\; x_t > 0 \; .
\end{equation}
Here $K_n$ is the modified Bessel function of order $n$, $\phi$ is
the external flux, and $\lambda=1/\mu$ is the London penetration length. 
This field shape in Eq.~(\ref{London}) is acceptable provided that
$\lambda \gg \xi$, $\xi$ being the coherence length which represents the
typical size scale of the density variations of the magnetic
monopole condensate (the dual version of the Cooper condensate).
Due to this condition, the solution given in~(\ref{London}) is appropriate
only for superconductors that, in the language of ordinary superconductivity,
are classified as type-II superconductors. However, within the dual superconductor model it is expected that, having the source and sink of the color fields in the QCD vacuum,
tube-like structures arise irrespective of the value of the $\lambda/\xi$ ratio.
The main flaw of the ansatz in Eq.~\eqref{London} is, instead, the divergence of the field value at $x_t = 0$.
In this respect a more adequate solution was constructed long ago in 
Ref.~\cite{Clem:1975aa}, where, starting from a simple variational
model for the magnitude of the normalized order parameter of an isolated vortex,
an analytic expression was derived for both the magnetic field and supercurrent density,
that solves Amp\`ere's law and the Ginzburg-Landau equations.
Only recently it was suggested and successfully
adopted~\cite{Cea:2012qw,Cea:2013oba,Cea:2014uja,Cea:2014hma} in order to describe
the transverse distribution of the chromoelectric flux tube:
\begin{equation}
\label{clem1}
E_l(x_t) = \frac{\phi}{2 \pi} \frac{1}{\lambda \xi_v} \frac{K_0(R/\lambda)}
{K_1(\xi_v/\lambda)} \; ,
\end{equation}
where
\begin{equation}
\label{rrr}
 R=\sqrt{x_t^2+\xi_v^2}
\end{equation}
and $\xi_v$ is a variational core-radius parameter.
Equation~(\ref{clem1}) can be rewritten as
\begin{equation}
\label{clem2}
E_l(x_t) =  \frac{\phi}{2 \pi} \frac{\mu^2}{\alpha} \frac{K_0[(\mu^2 x_t^2 
+ \alpha^2)^{1/2}]}{K_1[\alpha]} \; ,
\end{equation}
with
\begin{equation}
\label{alpha}
\mu= \frac{1}{\lambda} \,, \quad \frac{1}{\alpha} =  \frac{\lambda}{\xi_v} \,.
\end{equation}
By fitting Eq.~(\ref{clem2}) to flux-tube data, one can get 
both the penetration length $\lambda$ and the ratio of the penetration length 
to the variational core-radius parameter, $\lambda/\xi_v$. Moreover,   
the Ginzburg-Landau $\kappa$ parameter, which in ordinary superconductivity discriminates the type of superconductor, can be obtained by
\begin{equation}
\label{landaukappa}
\kappa = \frac{\lambda}{\xi} =  \frac{\sqrt{2}}{\alpha} 
\left[ 1 - K_0^2(\alpha) / K_1^2(\alpha) \right]^{1/2} \,,
\end{equation}
whereas the coherence length $\xi$ can be determined by combining  
Eqs.~(\ref{alpha}) and~(\ref{landaukappa}).
We will consider two more observables which give information about the
structure and the properties of the chromoelectric flux tube:
the {\em mean square root width},
\begin{equation}
\label{width}
\sqrt{w^2} = \sqrt{\frac{\int d^2x_t \, x_t^2 E_l(x_t)}{\int d^2x_t \, E_l(x_t)}}
= \sqrt{\frac{2 \alpha}{\mu^2} \frac{K_2(\alpha)}{K_1(\alpha)}}
\end{equation}
and the square root of the {\em energy per unit length}, normalized to the
flux $\phi$,
\begin{equation}
\label{energy}
\frac{\sqrt{\varepsilon}}{\phi} = \frac{1}{\phi} \sqrt{ \int d^2x_t \,
  \frac{E_l^2(x_t)}{2} } =  \sqrt{ \frac{\mu^2}{8 \pi} \,
  \left(1-\left(\frac{K_0(\alpha)}{K_1(\alpha)}\right)^2\right)} \;.
\end{equation}
\section{Lattice setup and numerical results}
\label{setup}
We performed all simulations, both for pure gauge $SU(3)$ and (2+1)-flavor QCD,
on $32^4$ lattices, making use of the publicly available MILC
code~\cite{MILC}, suitably modified in order to introduce the relevant
observables. The typical statistics of each run was about 4000-5000;
to allow for thermalization we typically discarded a few thousand sweeps. The error
analysis was performed by the jackknife method over bins at different blocking levels.  

The lattice discretization that we used for the pure gauge $SU(3)$ is the
standard Wilson action, with the physical scale set assuming for the
string tension the standard value of $\sqrt{\sigma} = 420$~MeV and using
the parameterization~\cite{Edwards:1998xf}
\bea
\label{sqrt-sigma-SU3}
\left ( a \, \sqrt{\sigma} \right )(g) &=& f_{{\rm{SU(3)}}}(g^2) 
\left \{ 1+0.2731\,\hat{a}^2(g)  \right .\\
&-&0.01545\,\hat{a}^4(g) +0.01975\,\hat{a}^6(g) \left . \right \}/ 0.01364 \;  , \nonumber
\eea
\[
\hat{a}(g) = \frac{f_{{\rm{SU(3)}}}(g^2)}{f_{{\rm{SU(3)}}}(g^2(\beta=6))} 
\;, \;
\beta=\frac{6}{g^2} \,, \;\;\; 5.6 \leq \beta \leq 6.5\;,
\]
with
\beq
\label{fsun}
f_{{\rm{SU(3)}}}(g^2) = \left( {b_0 g^2}\right)^{- b_1/2b_0^2} 
\, \exp \left( - \frac{1}{2 b_0 g^2} \right) \;, \;\;\; b_0 \, = \, \frac{11}{(4\pi)^2} \; \; , \; \; b_1 \, = \, \frac{102}{(4\pi)^4} \,.
\eeq
The used value of the string tension comes from the universal slope of the
Regge trajectories and the phenomenology of heavy quark systems~\cite{PhysRevD.21.203}.
For (2+1)-flavor QCD we adopted the HISQ/tree action~\cite{Bazavov:2010ru}
and worked on the line of constant physics determined in~\cite{Bazavov:2011nk},
by adjusting the coupling and the bare quark masses so as to keep the strange quark
mass $m_s$ fixed at its physical value with the light-to-strange mass ratio
$m_l/m_s = 1/20$, corresponding to a pion mass of 160 MeV. The scale was set through
the slope of the static quark-antiquark potential evaluated on zero-temperature
lattices, using the results of Ref.~\cite{Bazavov:2011nk}.
\subsection{Smoothing procedure}
\label{smoothing}
The connected correlator defined in Eq.~(\ref{rhoW}) suffers from large
fluctuations at the scale of the lattice spacing, which are responsible
for a bad signal-to-noise ratio. To extract the physical information carried
by fluctuations at the physical scale (and, therefore, at large distances
in lattice units) we smoothed out configurations by the {\em smearing}
procedure. Our setup consisted of (just) one step of HYP
smearing~\cite{Hasenfratz:2001hp} on the temporal links, with smearing
parameters $(\alpha_1,\alpha_2,\alpha_3) = (1.0, 0.5, 0.5)$, and
$N_{\rm APE}$ steps of APE smearing~\cite{Falcioni1985624} on the spatial links, with 
smearing parameter $\alpha_{\rm APE} = 0.167$. Here $\alpha_{\rm APE}$ is the ratio 
between the weight of one staple and the weight of the original link.
The optimal number of smearing step was found by looking at the
smearing step at which our direct observable $E_l(x_t)$ showed the
largest signal-to-noise ratio, with the smearing parameter tuned in
such a way that in the $E_l(x_t)$  vs 'smearing step' plot we could see
a clear plateau.

\begin{figure}[htb] 
\centering
\includegraphics[scale=0.5,clip]{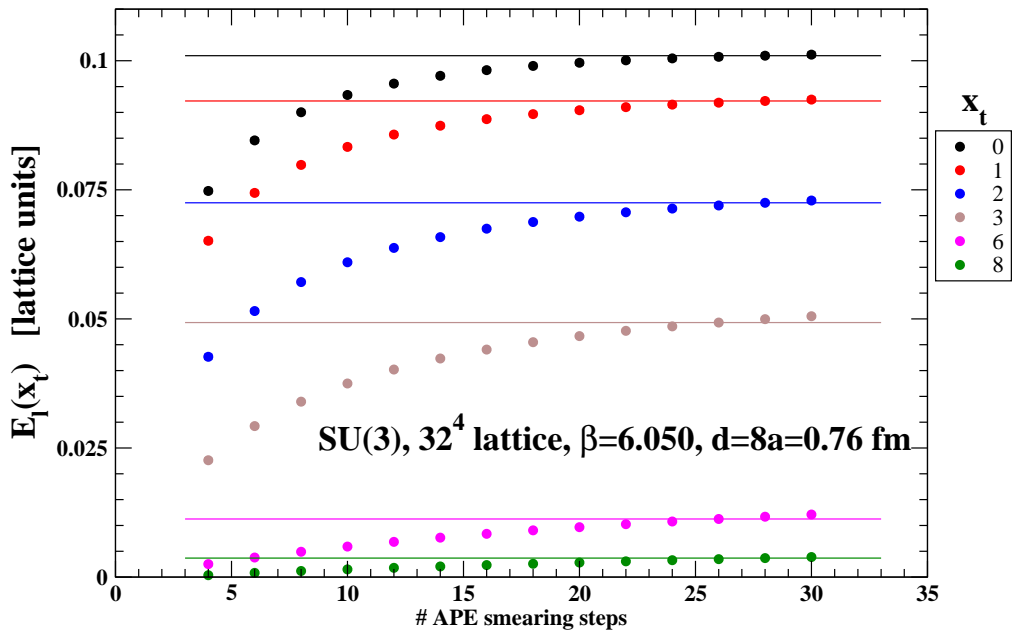}
\caption{(color online). Behavior of the longitudinal chromoelectric field
  $E_l$, on a given lattice and for various values of the distance from
  the axis connecting the static sources, {\it versus} the number of
  APE smearing steps on the spatial links.}
\label{fig:El_vs_smear}
\end{figure}

In Fig.~\ref{fig:El_vs_smear} we show the behavior under smearing of
the longitudinal chromoelectric field $E_l(x_t)$ in pure gauge $SU(3)$ for a
physical distance between the static sources equal to 0.76~fm. We can
see that, for each value of the distance $x_t$ in the direction transverse
to the axis connecting the sources, a clear plateau is reached after a
sufficiently large number of smearing steps. 
A similar behavior was
observed in {\em all} simulations we performed, both in pure gauge $SU(3)$
and in (2+1)-flavor QCD. All results concerning the chromoelectric field
$E_l(x_t)$ presented in the following will always refer to determinations
on smeared configurations, after a number of smearing steps $N_{\rm APE}$
such that the plateau is reached for {\em all} considered values of $x_t$.
The typical value of $N_{\rm APE}$ ranges between 25 and 50.
\subsection{Continuum scaling}
\label{scaling}
Our aim is to determine the physical properties of the chromoelectric flux tube
in the {\em continuum}, for this reason, we have preliminarily checked
that our simulations are performed in a region of values of the coupling
$\beta$ where continuum scaling holds.

We have hence measured the longitudinal chromoelectric field generated when
the static sources are located at the {\em same} physical distance $d$, but
for two {\em different} values of the coupling $\beta$ or, equivalently,
of the distance in lattice units. This test was performed both in pure
gauge $SU(3)$ and in (2+1)-flavor QCD.

\begin{figure}[htb]
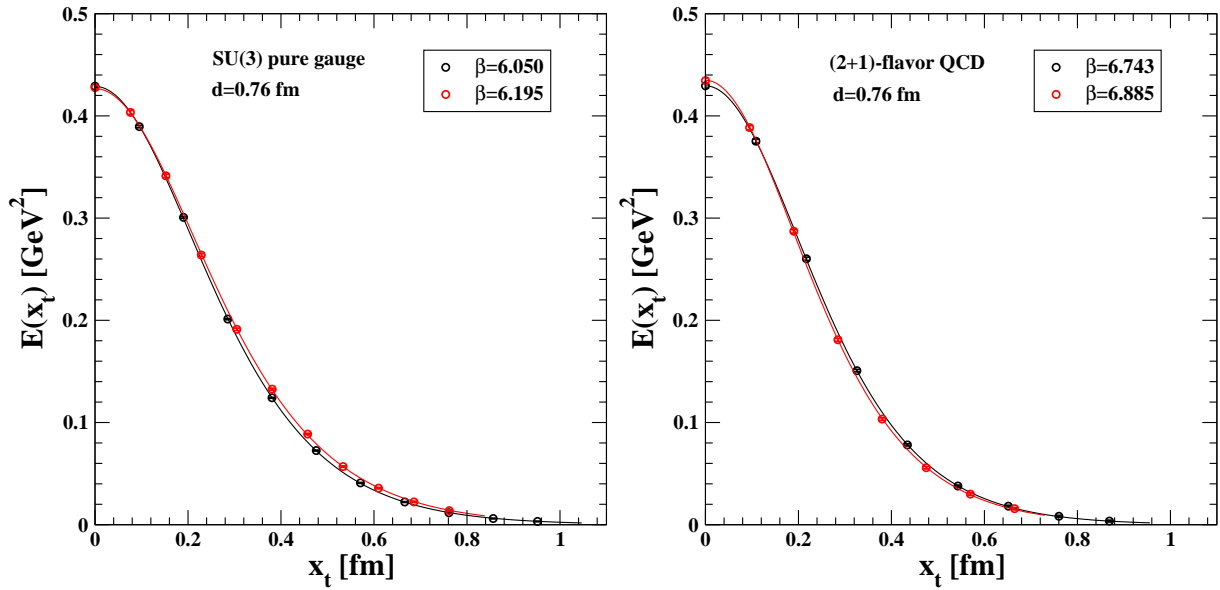
 
\centering
\includegraphics[scale=0.45,clip]{E_physical_scaling_SU3.eps}
\includegraphics[scale=0.45,clip]{E_physical_scaling_QCD.eps}
\caption{(color online). Behavior of the longitudinal chromoelectric field
  $E_l$ (in physical units) {\it versus} the distance $x_t$ (in physical units)
  from the axis connecting the static sources, at the {\em same} physical
  distance between the sources as obtained for two {\em different} $\beta$
  values, in the case of the $SU(3)$ pure gauge theory (left panel) and
  of (2+1)-flavor QCD (right panel).}
\label{fig:scaling}
\end{figure}

In Fig.~\ref{fig:scaling} we present the outcome of this test: in the left
panel we show the (smeared) chromoelectric field in pure gauge $SU(3)$
{\it versus} the transverse distance $x_t$ in physical units, when the sources
are placed at distance $8a$ and $10a$ ($a$ is the lattice spacing) at
$\beta=6.050$ and $\beta=6.195$, respectively, so that, according to
Eq.~(\ref{sqrt-sigma-SU3}), the physical distance is, in both cases, equal
to 0.76~fm in physical units. The right panel of Fig.~\ref{fig:scaling}
shows the result of a similar analysis in (2+1)-flavor QCD: here the
distances in lattice units were fixed at distances $7a$ and $8a$, for
$\beta=6.743$ and $\beta=6.885$, respectively, so that the physical distance
$d$ between the sources is again equal to 0.76~fm. In both cases an
almost perfect scaling can be observed, thus making us confident that,
for the observable of interest in this work, the continuum scaling is
reached in $SU(3)$ (at least) for $\beta=6.050$ and in (2+1)-flavor QCD
(at least) for $\beta=6.743$. Another hint from the results shown in 
Fig.~\ref{fig:scaling} is that our smearing procedure is robust: had the
smearing procedure badly corrupted the physical signal for the chromoelectric
field, it would have been quite unlikely to obtain such a nice scaling.
\subsection{Shape of the flux tube}
\label{results}
\begin{figure}[tb]
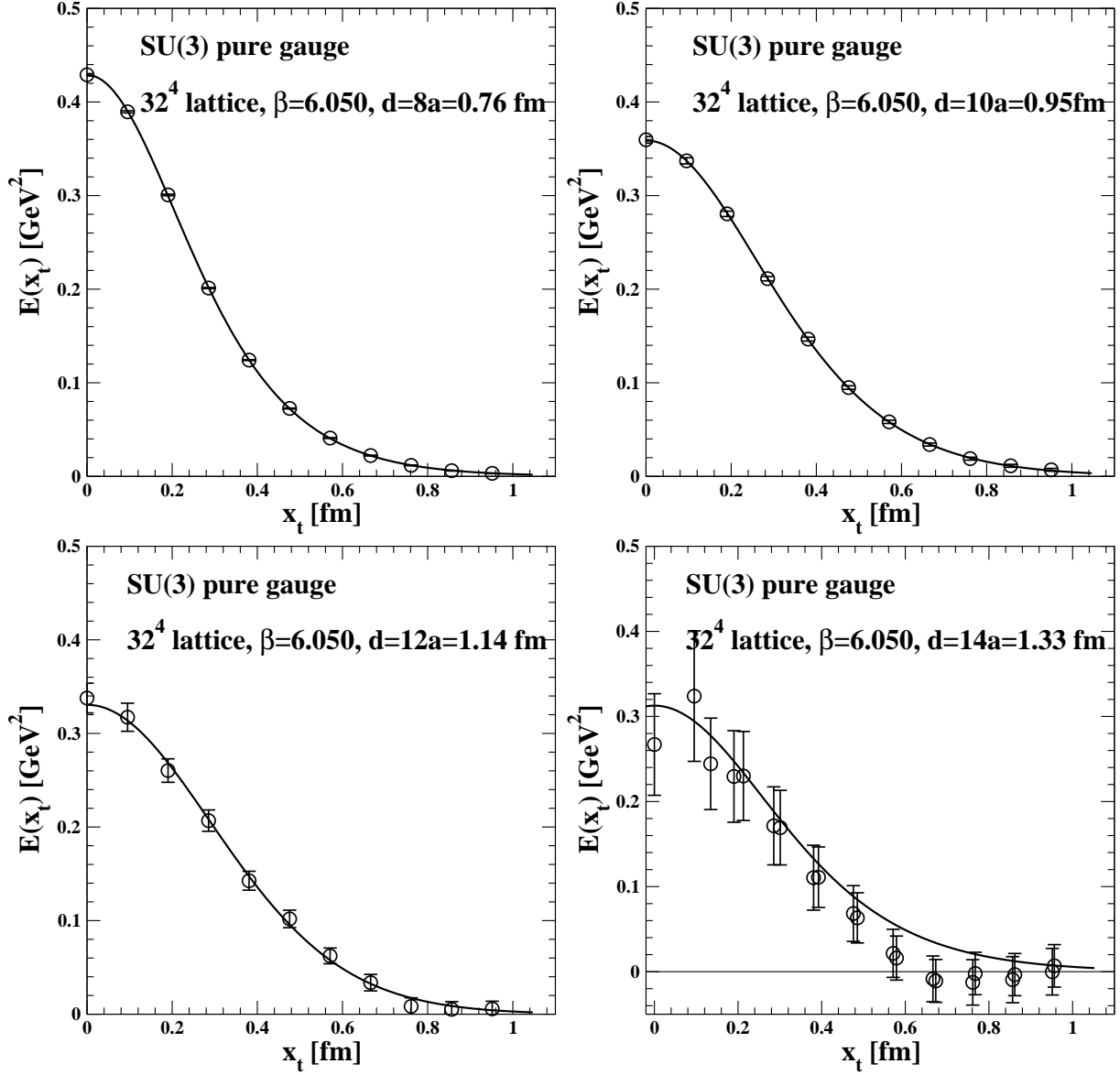
 
\centering
\includegraphics[scale=0.45,clip]{E_physical_SU3_dist8.eps}
\includegraphics[scale=0.45,clip]{E_physical_SU3_dist10.eps}

\includegraphics[scale=0.45,clip]{E_physical_SU3_dist12.eps}
\includegraphics[scale=0.45,clip]{E_physical_SU3_dist14.eps}
\caption{Behavior of the longitudinal chromoelectric field
  $E_l$ (in physical units) {\it versus} the distance $x_t$ (in physical units)
  from the axis connecting the static sources, at four different values
  of the physical distance between the sources in the $SU(3)$ pure gauge theory.}
\label{fig:SU(3)_distance}
\end{figure}
\begin{table}[h]
\begin{center} 
\caption{Flux tube parameters for the $SU(3)$ pure gauge theory at various
distances between the static sources.}
\label{tab:fit_summary_SU3}
\footnotesize
\begin{tabular}{|c|c|c|c|c|c|c|c|c|c|c|}
\hline\hline
$\beta$ & $d$ [fm] & $\phi$ & $\lambda$ [fm]& $\kappa=\lambda/\xi$ & $\xi$ [fm] & $\sqrt{w^2}$ [fm]& $\sqrt{\varepsilon}/\phi$ [GeV]\\ \hline
6.050 &0.76 &5.143(39) &0.164(5) &0.348(208) &0.472(283) &0.458(17) &0.133(5) \\
6.195 &0.76 &4.862(40) &0.155(6) &0.306(167) &0.506(278) &0.443(19) &0.137(6) \\
6.050 &0.95 &5.287(109)&0.146(17)&0.170(63)  &0.859(331) &0.479(69) &0.123(19)\\
6.050 &1.14 &5.218(371)&0.143(40)&0.145(48)  &0.983(428) &0.488(140)&0.120(35)\\
6.050 &1.33 &5.000(292)&0.169(16)&0.236(109) &0.715(335) &0.512(114)&0.117(30)\\
\hline\hline 
\end{tabular} 
\end{center}
\end{table}
We determined, both in $SU(3)$ and in (2+1)-flavor QCD, the dependence of the
longitudinal chromoelectric field $E_l$ on the transverse distance $x_t$,
by Monte Carlo evaluations of the expectation value of the operator
$\rho_{W,\mu\nu}^{\rm conn}$ (see Eq.~(\ref{rhoW})) over smeared ensembles, and
compared it with the function given in Eq.~(\ref{clem2}).
Such comparison was carried out for a few values of the distance $d$ between the
static sources, at values of the $\beta$-coupling lying inside the
continuum scaling region.

In Fig.~\ref{fig:SU(3)_distance} we report the results in physical units
of our simulations for the case of the $SU(3)$ pure gauge theory: we can see
that, in an interval of distances between the sources ranging from 0.76~fm
to 1.33~fm, data for $E_l(x_t)$ are nicely fitted by the function given
in Eq.~(\ref{clem2}), with $\chi^2/{\mathrm{dof}}=\mathcal{O}(1)$.
For larger values of $d$ the statistical noise becomes
overwhelming, preventing us from extracting reliable estimates using the present setup.

The fundamental fit parameters $\lambda$ and $\xi$ are summarized in
Table~\ref{tab:fit_summary_SU3}, together with $\kappa=\lambda/\xi$, the
mean square root width, $\sqrt{w^2}$, and the square root of the
normalized energy per unit length, $\frac{\sqrt{\varepsilon}}{\phi}$.
We can argue that the penetration length $\lambda$ is stable within errors
under variations of the distance between the sources, while there seems to
be a slow increase of $\xi$ and $\sqrt{w^2}$ as $d$ grows.
We notice also that the values of $\lambda$ obtained here for the case
of the $SU(3)$ pure gauge theory nicely compare with our previous 
determination~\cite{Cea:2013oba}, obtained for a distance $d$=0.62~fm on
a smaller lattice and using a slightly different setup for the smearing
procedure. We find, on the other hand, that our values for $\sqrt{w^2}$ are a
bit larger than those found for an analogous observable in the very recent
analysis of Ref.~\cite{Bicudo:2017uyy}, where the flux tube
profile was determined through the {\em disconnected} plaquette-Wilson
loop correlator and an ansatz different from ours was used to interpolate it.
The relatively large error bars in our determinations for $\sqrt{w^2}$
do not allow us to make any firm statement about the possible nature
of the flux-tube widening.

\begin{figure}[h]
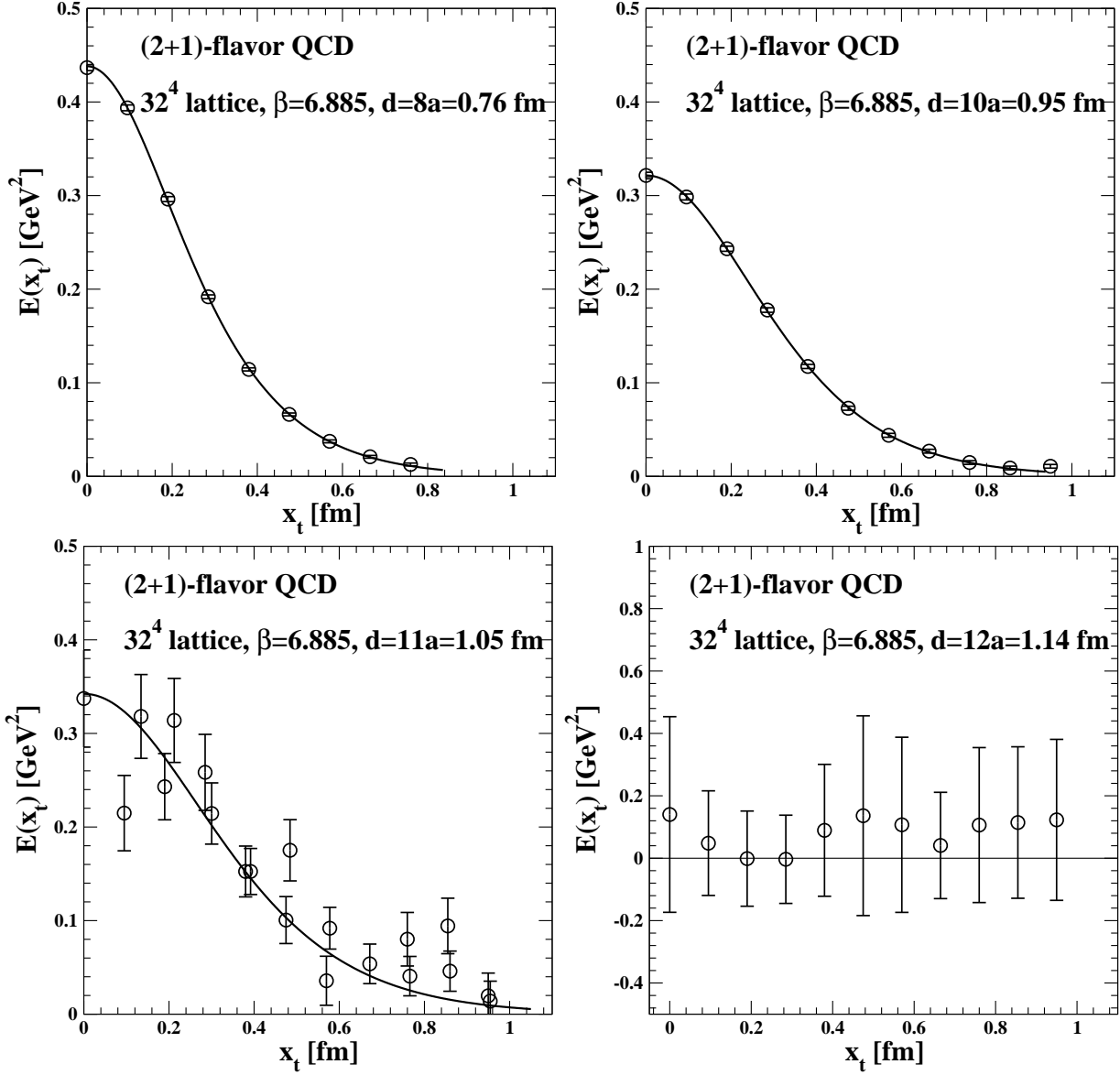
 
\centering
\includegraphics[scale=0.45,clip]{E_physical_QCD_dist8.eps}
\includegraphics[scale=0.45,clip]{E_physical_QCD_dist10.eps}

\includegraphics[scale=0.45,clip]{E_physical_QCD_dist11.eps}
\includegraphics[scale=0.45,clip]{E_physical_QCD_dist12.eps}
\caption{Behavior of the longitudinal chromoelectric field
$E_l$ (in physical units) {\it versus} the distance $x_t$ (in physical units)
from the axis connecting the static sources, at four different values
of the physical distance between the sources in (2+1)-flavor QCD.}
\label{fig:QCD_distance}
\end{figure}
%

A similar analysis was performed in (2+1)-flavor QCD, where the values of $d$
were taken in the range 0.76~fm to 1.14~fm. Results are presented
in Fig.~\ref{fig:QCD_distance} and in Table~\ref{tab:fit_summary_QCD}.
We observe first of all that at $d$=0.76~fm the value of
$\lambda$ obtained in (2+1)-flavor QCD is fairly consistent with the one
obtained in $SU(3)$. However, comparing the results at $d$=0.76~fm and
at $d$=0.95~fm, there is an indication that $\sqrt{w^2}$ keeps constant,
whereas $\lambda$ increases and $\xi$ decreases. The determinations at the
latter distance, however, are plagued by large uncertainties and should be
handled with care.

Another interesting fact is that at distance $d$=1.14~fm, the chromoelectric
longitudinal field $E_l(x_t)$ seems to fluctuate around zero, although
within large error bars, while, at the same distance, a clear nonzero
signal for $E_l(x_t)$ could be detected in the $SU(3)$ pure gauge theory.
This circumstance could be the consequence of larger statistical fluctuations
induced by dynamical fermions or be the signature of the phenomenon
of ``string breaking''~\cite{Bali:2005fu,Kratochvila:2002vm},
which could take place just around this distance~\cite{Bali:2005fu,Koch:2015qxr}.
\begin{table}[h]
\begin{center} 
\caption{Flux tube parameters for the (2+1)-flavor QCD at various
distances between the static sources.}
\label{tab:fit_summary_QCD}
\footnotesize
\begin{tabular}{|c|c|c|c|c|c|c|c|c|c|}
\hline\hline
$\beta$ & $d$ [fm] & $\phi$ & $\lambda$ [fm]& $\kappa=\lambda/\xi$ & $\xi$ [fm] & $\sqrt{w^2}$ [fm]& $\sqrt{\varepsilon}/\phi$ [GeV]\\ \hline
6.743 &0.76 &4.431(57) &0.141(8) &0.272(137) &0.521(265) &0.415(29) &0.145(11)\\
6.885 &0.76 &4.331(82) &0.155(11)&0.390(252) &0.398(259) &0.423(34) &0.145(29)\\
6.885 &0.95 &4.272(131)&0.154(21)&0.236(108) &0.653(312) &0.527(50) &0.128(22)\\
6.885 &1.05 &5.580(441)&0.174(14)&0.236(109) &0.736(343) &0.527(50) &0.113(11)\\
\hline\hline 
\end{tabular} 
\end{center}
\end{table}

The energy per unit length $\epsilon$, given in the last columns of
Tables~1 and~2, represents the contribution of the chromoelectric field to
the string tension, under the hypothesis of a uniform field along the axis
of the flux tube; it turns out to be of the same order of the measured string
tension, but cannot be directly compared with it, since the latter includes
also the contribution from the vacuum energy and the (negligible)
contribution from the other field components.

\section{Discussion}
\label{discussion}
In this paper we have studied the chromoelectric field in the direction
longitudinal to the line connecting a static quark-antiquark pair,
and its shape on the transverse plane cutting this line in its middle point.
This investigation has been performed both in the $SU(3)$ pure gauge theory and
in (2+1)-flavor QCD, with the aim of assessing any possible effect from the
variation of the physical distance between the static sources.
We considered distances extending in the range 0.76~fm to 1.33~fm.

Let us first summarize the common features we observed in the two theories:
\begin{itemize}
\item the transverse shape of the longitudinal chromoelectric field midway
  between the sources is accessible to Monte Carlo simulations through the
  measurement of the expectation value of a suitable connected operator
  (see Eq.~\eqref{rhoW}) on {\em smeared} configuration ensembles;

\item up to distances between the sources of about 1.5~fm in the case of the
  $SU(3)$ pure gauge theory and of about 1.1~fm in (2+1)-flavor QCD, this
  transverse shape is nicely described by the function~(\ref{clem2}),
  which is the dual version of a solution derived long ago in
  Ref.~\cite{Clem:1975aa} for the magnetic field generated by a single vortex
  inside an ordinary superconductor;

\item the values of the parameters entering this function, namely the (dual
  versions of) the London penetration length $\lambda$ and the coherence
  length $\xi$, as extracted from the fit to numerical data, indicate that
  the vacuum behaves as a type-I superconductor.
\end{itemize}

In the specific case of the $SU(3)$ pure gauge theory, we found that
the parameter $\lambda$ and the mean width of the field transverse profile
remain fairly constant under variation of the distance between the
sources, whereas $\xi$ shows a tendency to increase with the distance, though
within large uncertainties. The stability of $\lambda$ supports the validity
of the dual superconductivity model over the considered range of distances
between the sources. It would be interesting to refine the numerical techniques
and to check how far sources must be located to observe the break-up of
the dual superconductivity picture and the onset of the effective string
description (see~\cite{Baker:2015zlm} for a nice discussion about the
interplay between the two pictures).

In (2+1)-flavor QCD the scenario is less clean: due to the larger
uncertainties and/or to the possible insurgence of new phenomena, such
as the ``string breaking'', the range of distances we could explore
is smaller than in the pure gauge theory and it is thus more difficult
to identify a clear trend in the values of the parameter of the transverse
field profile. Data seem to suggest that $\lambda$ increases with the
distance between the sources, whereas the mean width remains stable. However,
further investigations and more efficient algorithms are needed to achieve
firmer conclusions.
\section*{Acknowledgments}
We thank Marshall Baker for useful comments on this work.
This investigation was in part based on the MILC collaboration's public
lattice gauge theory code. See
{\url{http://physics.utah.edu/~detar/milc.html}}.
Numerical calculations have been made possible through a CINECA-INFN
agreement, providing access to resources on GALILEO and MARCONI at CINECA.


%

\end{document}